\documentclass[aps,prb,
12pt,
]{revtex4-1}
\usepackage{bm}
\usepackage{amsmath}    
\usepackage{amssymb}    
\usepackage{graphicx}   
\usepackage{booktabs}
\usepackage{xcolor}
\usepackage[hang]{subfigure}
\usepackage{hyperref}
\DeclareMathOperator{\sech}{sech}

\DeclareMathOperator{\arctanh}{arcTanh}
\newcommand{\jun}{junction }
\newcommand{\juns}{junctions }
\newcommand{\Jos}{Josephson }

\newcommand{\ann}{annular }
\newcommand{\conf}{confocal }

\begin{document}
\title[Roberto Monaco]{Field Cooled Annular Josephson Tunnel Junctions}
\thanks{Monaco et al 2020 Supercond. Sci. Technol. https://doi.org/10.1088/1361-6668/ab92e7}
\author{Roberto Monaco}
\email[Corresponding author e-mail address:]{  r.monaco@isasi.cnr.it and roberto.monaco@cnr.it}
\affiliation{CNR-ISASI, Institute of Applied Sciences and Intelligent Systems ''E. Caianello'', Comprensorio Olivetti, 80078 Pozzuoli, Italy and\\ International Institute for Advanced Scientific Studies (IIASS), Vietri sul Mare, Italy}

\author{Jesper Mygind}
\affiliation{DTU Physics, B307, Technical University of Denmark, DK-2800 Lyngby, Denmark}
\email{myg@fysik.dtu.dk}

\author{Valery P. Koshelets}
\affiliation{Kotel'nikov Institute of Radio Engineering and Electronics,
Russian Academy of Science, Mokhovaya 11, Bldg 7, 125009 Moscow, Russia.}
\email{valery@hitech.cplire.ru}


\begin{abstract}

\begin{footnotesize}
We investigate the physics of planar annular Josephson tunnel junctions quenched through their transition temperature in the presence of an external magnetic field. Experiments carried out with long $Nb/Al$-$AlOx/Nb$ annular junctions showed that the magnetic flux trapped in the high-quality doubly-connected superconducting electrodes forming the junction generates a persistent current whose associated magnetic field affects the both the static and dynamics properties of the junctions. More specifically, the field trapped in the hole of one electrode combined with a d.c. bias current induces a viscous flow of dense trains of Josephson vortices which manifests itself through the sequential appearance of displaced linear slopes, Fiske step staircases and Eck steps in the junction's current-voltage characteristic. Furthermore, a field shift is observed in the first lobe of the magnetic diffraction pattern. The effects of the persistent current can be mitigated or even canceled by an external magnetic field perpendicular to the junction plane. The radial field associated with the persistent current can be accurately modeled with the classical phenomenological sine-Gordon model for extended one-dimensional Josephson junctions. Extensive numerical simulations were carried out to disclose the basic flux-flow mechanism responsible for the appearance of the magnetically induced steps and to elucidate the role of geometrical parameters. It was found that the imprint of the field cooling is enhanced in confocal annular junctions which are the natural generalization of the well studied circular annular junctions.      
\end{footnotesize}
\noindent \textcolor{blue}{\url{https://doi.org/10.1088/1361-6668/ab92e7}} 


\end{abstract}

\maketitle
\section{Introduction}

There is a continuously growing interest for novel applications and multi-fluxon dynamic states in annular \Jos systems \cite{Lee19, WM19,Rahmonov20}. Recently \cite{PRB19}, the unidirectional collective motion of a dense train of fluxons in Josephson junctions, called \textit{Josephson flux-flow}, has been first reported in current-biased planar Annular Josephson Tunnel Junctions (AJTJs) under the application of an in-plane uniform magnetic field generating \textit{flux-flow steps} (FFSs) in their current-voltage characteristics. More specifically, FFSs carrying a large supercurrent, which gauges the robustness of the flux-flow state, have been experimentally observed and numerically reproduced only in the so-called \textit{\conf} AJTJs in which the internal and external boundaries of the annular tunnel barrier are closely spaced confocal ellipses \cite{JLTP16b,JPCM16}, rather than concentric circles as for in the classical \textit{circular} AJTJs. The physics of Josephson tunnel junctions is known to drastically depend on their geometrical configurations \cite{Barone}; indeed, the phenomenology of a \conf AJTJ is strongly affected by its aspect ratio, $\rho$, defined as the ratio of the mean length of the minor axes to the mean length of the major axes of the annulus \cite{JLTP18}. Large magnetically induced steps were observed in \conf AJTJs with large aspect ratio provided that the in-plane uniform magnetic field is applied perpendicular to the junction major axis. As in linear one-dimensional Josephson tunnel junctions \cite{Nagatsuma83}, the voltage of the FFS increases nearly linearly with the strength of the externally applied in-plane magnetic field above a threshold value, called the \textit{critical field}, needed to first suppress the junction zero-voltage critical current. 
\noindent An alternative way to modulate the supercurrent of a planar \Jos tunnel \jun is to apply a transverse magnetic field, i.e., perpendicular to the \jun plane. The field lines bend around the specimen that is in the Meissner state and the field induces shielding currents in its electrodes \cite{rc,hf,miller}. In turn, the demagnetizing currents generate a local magnetic field with a component threading the \Jos barrier. The result of a transverse field strongly depends on the geometry of the electrodes and on how close to the barrier the shielding currents circulate. These effects have been investigated both theoretically and experimentally for rectangular as well as for \ann \juns \cite{JAP07,JAP08,PRB09}; furthermore, it has been demonstrated that for AJTJs made by specularly symmetric electrodes a transverse magnetic field is equivalent to an in-plane field applied in the direction of the current flow. The transverse critical field is much smaller than its in-plane analog \cite{SUST15,SUST18}. It is therefore not surprising that the flux-flow state can be established in \conf AJTJs also by a applying a transverse magnetic field and the resulting FFSs are indistinguishable from those induced by an in-plane magnetic field. Yet another way may exist to obtain a Josephson flux-flow in a AJTJ that does not require the application on an external magnetic field. It might exploit the permanent magnetic flux (strictly fluxoid) trapped in hole of a doubly connected electrode of a AJTJ when the phase transition from the normal to the superconducting state is carried out in a sufficiently large magnetic field; this procedure is commonly called \textit{field cooling} (FC). Due to the superconducting wave-function only having a single value, the fluxoid can only exist in quantized units and is time-independent, i.e., it is conserved when the cooling field is removed once the cool down is completed. Provided that at least one of the junction's electrodes is doubly connected, the permanent currents that circulate to maintain the trapped fluxoid can be large enough to induce and sustain the flow of Josephson vortices, even in the absence of any externally applied magnetic field. Discussions of flux trapping in superconducting thin films are almost as old as Josephson junction technologies \cite{vanDuzer} and trapping of residual or stray magnetic fields degrades and, in extreme cases, destroys the performance of Josephson devices and constitutes the most serious limitation to the integration of superconducting digital circuits \cite{Polyakov07}. 

\noindent In AJTJs, at variance, the quench in a transverse field can be used to our advantage for the creation of a permanent magnetic field. The purpose of this work is to investigate the effects of the FC on AJTJs and to provide an overall insight on the different trapping phenomena occurring during a quench. \textcolor{black}{It will be shown that the passive magnetic field generated by the persistent current can efficiently replace the external field induced by coils, solenoids or control lines.} 

The paper is organized into four sections. Section II describes the experimental findings obtained with low-loss $Nb/Al$-$AlOx/Nb$ window-type \conf AJTJs; we first illustrate the consequences of a transverse magnetic field applied to samples cooled in the absence of any external field and then present the same data with the \ann \juns cooled in transverse magnetic induction fields of different strength in the high microtesla range. In Sec.III we review the theoretical modeling of a current-biased AJTJ subjected to an external magnetic field in the framework of a modified and perturbed sine-Gordon equation; we then extend the model to take into account the magnetic field induced by permanent circulating currents and present numerically calculated current-voltage characteristics (IVCs) with parameters taken from the experiments that describe the dynamical state in the flux-flow regime. The numerical results are compared with experiment, and an overall good agreement is found. Some comments and the conclusions of our work are presented in Sec. IV.

\section{The measurements}

\subsection{The samples and the experimental setup}

\begin{figure}[b]
\centering
\includegraphics[width=7.5cm]{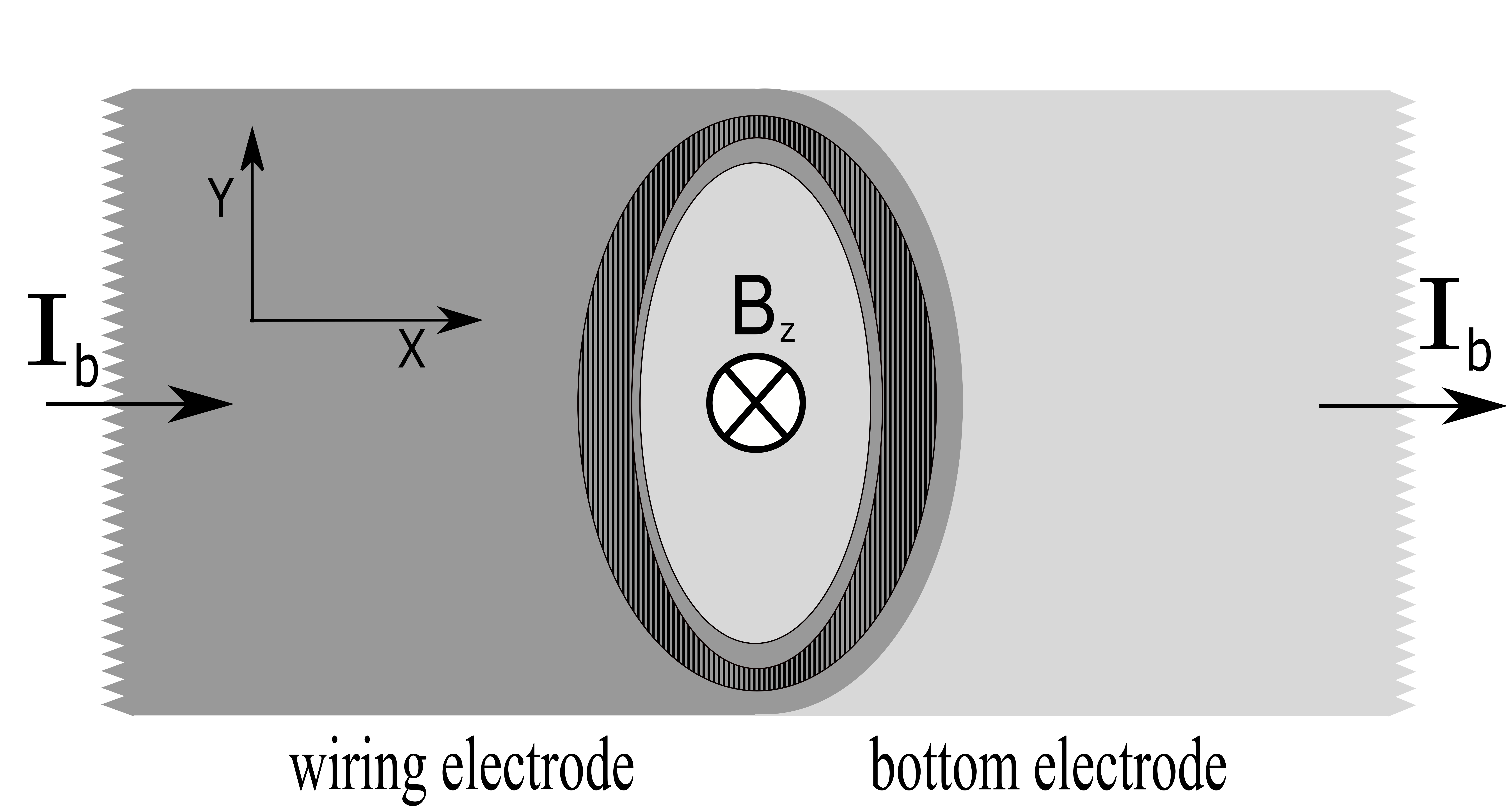}
\caption{Drawing of a Lyngby-type \textit{confocal} annular Josephson tunnel junction laying in the $X$-$Y$ plane. The dark area, delimited by two closely spaced ellipses having the same foci, represents the junction tunneling area. As the eccentricity of the ellipses vanishes, the confocal annulus progressively reduces to a circularly symmetric annulus (with uniform width). The dc bias current, $I_b$, flows in the two horizontal electrodes. The base electrode (light gray) is simply connected, while the top/wiring electrode (dark gray) is doubly connected, i.e., has a hole. A transverse magnetic field, $B_z$, can be applied perpendicular to the junction's plane by means of a superconducting cylindrical coil with its axis oriented along the $Z$-direction.}
\label{ConfAnn}
\end{figure}

\begin{table*}[b]
	\centering
\scriptsize
\begin{tabular}{lc}\hline
{\it $\qquad \qquad \qquad$ Geometrical details} & \\  \hline
Aspect ratio, $\rho$ & 1/4\\ \hline
Interfocal distance, $2c$ & 90.2\,$\mu$m\\ \hline
Minimum width, $\Delta w_{min}$ & 2.1\,$\mu$m\\ \hline
Maximum width, $\Delta w_{max}$ & 8.4\,$\mu$m\\ \hline
Mean perimeter, $L$ & 200\,$\mu$m\\ \hline
Area, $A$ & 1310$\,\mu$m$^2$\\ \hline
$\bar{\nu}\equiv \arctanh \rho$ & 0.26\\ \hline
$\Delta \nu\equiv\Delta w_{min}/c \sinh \bar{\nu}$ & 0.18\\ \hline
{\it $\qquad \qquad \qquad$ Electrical parameters} & \\ \hline
Critical current density, $J_c$ & 4.7\,kA/cm$^2$\\ \hline
Josephson length, $\lambda_J$ & 3.9$\,\mu$m\\ \hline
Normalized length, $L/\lambda_J$ & $\approx$ 50\\ \hline
Maximum critical current, $I_c^{max}$ & 28\,mA\\ \hline
Gap quasiparticle current step, $\Delta I_g$ & 96\,mA\\ \hline
Subgap leakage current, $I_{sg}(2mV)$ & 4.6\,mA\\ \hline
$2\Delta$ gap voltage, $V_g$ & 2.85\,mV\\ \hline
\end{tabular}
\caption{Geometrical details of the tunneling area and electrical parameters (measured at $4.2\,K$) of the selected \conf \ann Josephson tunnel junction.}
		 \label{tableI}
\end{table*}


\noindent \textcolor{black}{In the experiments, we used high quality $Nb/Al$-$AlOx/Nb$ AJTJs fabricated on silicon substrates using the tri-layer technique in which the Josephson junction is realized in a window opened in an insulator layer. The nominal thicknesses of the bottom and top sputtered electrodes of the trilayer were, respectively, $190\,$nm and $65\, $nm. The junctions were patterned from the $Nb/Al$-$AlOx/Nb$ tri-layer by the reactive ion etching of the top $Nb$ layer using $CF_4$ (the $Al$-$AlOx$ bi-layer serves as etch stop layer) followed by a light wet anodization. The dielectric layer for junction insulation consists of a $200\,nm$-thick $SiO_2$ film, defined in a self-aligned lift-off procedure. The electric contact to the top electrode was realized by sputtering a $470\, $nm thick $Nb$ wiring layer having a residual resistivity ratio as large as $100$ and a critical temperature $T_c\cong 9.1\,K$.}

\noindent High-quality window-type $Nb/Al$-$AlOx/Nb$ \conf AJTJs were used for our investigation. The fabrication process and the geometrical layout can be found elsewhere \cite{VPK,Filippenko,SUST18,PRB19}. All our samples were designed with the so-called \textit{Lyngby-type} geometry\cite{davidson85} that refers to a specularly symmetric configuration in which the width of the current carrying electrodes matches one of the ellipse outer axis. One example of this geometry is sketched in Fig.~\ref{ConfAnn} where the dark area delimited by two closely spaced ellipses having the same foci represents the junction tunneling area. In this specific example the system's aspect ratio is $1/2$ that implies that the equatorial annulus width is twice the polar width. As the eccentricity of the ellipses vanishes, the confocal annulus progressively reduces to a circularly symmetric annulus (with uniform width). The dc bias current, $I_b$, flows parallel to the minor axis of the confocal annulus. The base electrode (light gray) is simply connected, while the top/wiring electrode (dark gray) is doubly connected, i.e., a quantized magnetic flux can be trapped in its elliptical hole. 

\noindent The experiments were done in an rf-shielded room immersing a cryoprobe in a liquid helium cryostat. The $3\times4.2mm^2$ $Si$chip was hold in the center of a long superconducting cylindrical solenoid whose axis was along the vertical direction to provide an in-plane magnetic field. In addition a transverse magnetic field, $B_z$, was applied by means of a superconducting cylindrical coil with its axis oriented along the $Z$-direction, i.e., perpendicular to the junctions plane. \textcolor{black}{The large magnetic sensitivity of long JTJs requires a careful shield of the Earth's magnetic field. Therefore, the chip-holder was magnetically shielded by means of two concentric superconducting $Pb$ cans surrounded by a long vacuum-tight cryoperm can. The chip holder, the superconducting shields, the cryoperm can and the coils were cooled all together down to $4.2\, K$ where all measurements were carried out. In the absence of of any externally applied magnetic field, several zero-field steps were observed in the low-voltage region of junctions I-V characteristic, indicating that the estimated residual magnetic field amounts to no more than few $\mu T$.}

\noindent A large number of \conf AJTJs were investigated having different geometrical and electrical parameters and all showed highly hysteretic IVCs with low subgap leakage currents, $I_{sg}$, compared to the current jump, $\Delta I_g$, at the $2\Delta$ gap voltage, $V_g$. Nominally identical samples made within the same fabrication run gave qualitatively similar results; therefore, the findings presented in this work pertain to just a representative one (for which the experimental data are more exhaustive). The geometrical details of the tunneling area for the selected \conf AJTJ and its relevant electrical parameters (measured at $4.2\,K$) are listed in Table I. \textcolor{black}{The critical current density of our samples was measured on electrically small cross-type \juns realized in the same wafer on different chips. The value of the Josephson penetration depth $\lambda_J$ was calculated assuming a $Nb$ London penetration of $90\,nm$ \cite{broom,VPK} and taking into account the the effect of the lateral idle region \cite{JAP95,Ustinov}.}

It is important to keep in mind that the \ann \jun considered in this section has just one hole that, as depicted in Fig.~\ref{ConfAnn}, is realized in the top electrode. It is worth noting that for fabrication requirements this elliptical hole does not follow the inner boundary of the barrier area.  



\vskip -80pt
\subsection{Zero-Field Cooling}

\begin{figure}[b]
\centering
\subfigure[ ]{\includegraphics[width=8cm]{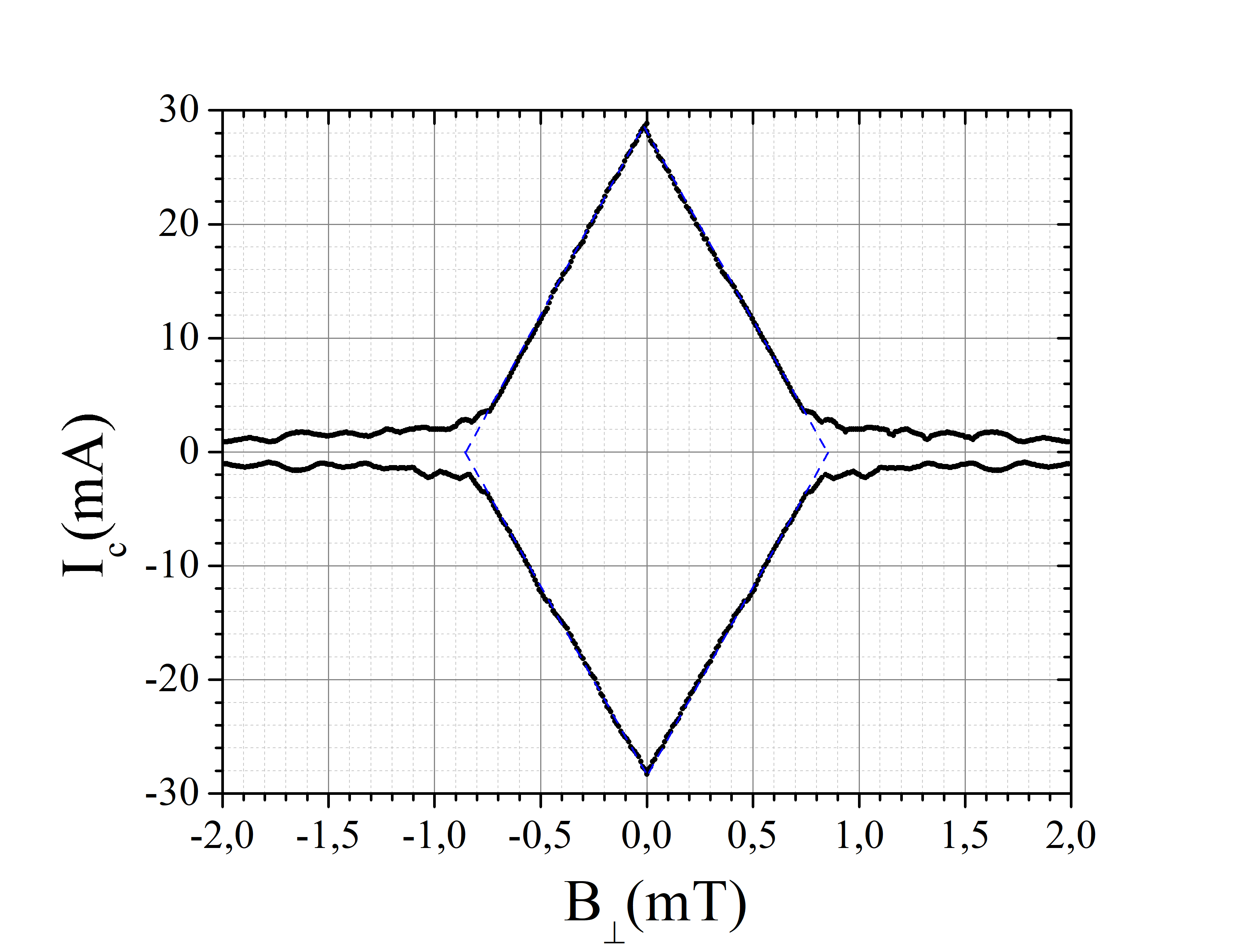}}
\subfigure[ ]{\includegraphics[width=8cm]{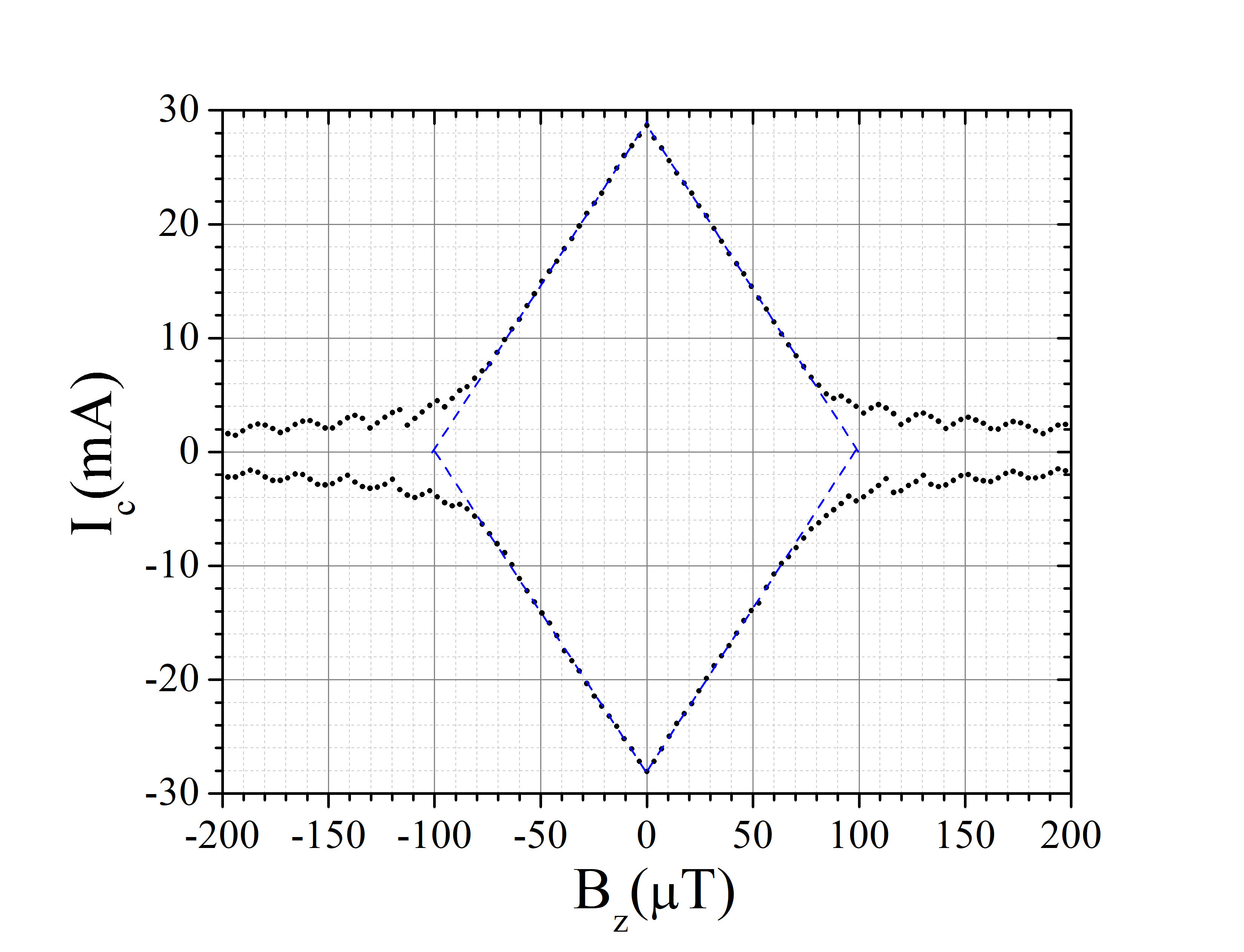}}
\caption{(Color online) Experimental magnetic diffraction patterns, $I_c(B)$, recorded at $4.2\,K$, after a zero-field cooling (ZFC) of the \conf AJTJ with different field orientations: (a) in-plane magnetic induction field, $B_{\bot}$, applied perpendicular to the annulus major diameter and (b) transverse magnetic induction  field, $B_z$, applied perpendicular to the junction's plane. The extrapolated dotted lines help to locate the critical fields.}
\label{ZFC_MDP}
\end{figure}

\noindent For the sake of clarity and completeness, we first report the experimental finding recorded at $T=4.2\,K$ for the sample cooled through its critical temperature in the absence of any external magnetic field.  On quenching the system from the normal to superconducting phase, causality prevents the \jun from adopting a uniform phase. This symmetry-breaking process, known as Kibble-Zurek mechanism \cite{kibble1,zurek1}, spontaneously generated one or more fluxons on a statistical basis \cite{PRB06,PRB08} with a probability that increases with the speed of the normal-to-superconducting transition; at the end of each zero-field quench the number of trapped fluxons is determined by inspecting the junction IVC and measuring the voltage of possible zero-field steps. Nevertheless, for the sake of simplicity, we will limit our attention to the cases in which no fluxon is trapped during the phase transition. Figures~\ref{ZFC_MDP}(a)-(b) display the magnetic diffraction patterns (MDP) of the zero-voltage critical current, $I_c(B)$, recorded at $4.2\,K$ after a zero-field cooling of the \conf AJTJ with, respectively, an in-plane magnetic induction field, $B_{\bot}$, perpendicular to the annulus major diameter \cite{note1} and a transverse magnetic induction field, $B_z$, perpendicular to the junction's plane (when a transverse field is applied, a circulating current is induced in the electrodes, but the magnetic flux up through the hole in the top electrode remains zero). It is seen that in both cases the main lobes of the MDPs show a linear dependence of the supercurrent, $I_c$, on the external field. The (first) critical fields are obtained by extrapolating to zero the MDP main lobe (see dotted lines): as expected, $B_z$ is almost one order of magnitude more efficient than $B_{\bot}$ to suppress the critical current. In both cases the applied fields are much smaller than the low-temperature Nb lower critical field, $B_{c1}^{Nb}\simeq 190\,mT$, that would drive the superconducting films into the mixed state.

\begin{figure}[b]
\centering
\includegraphics[width=9cm]{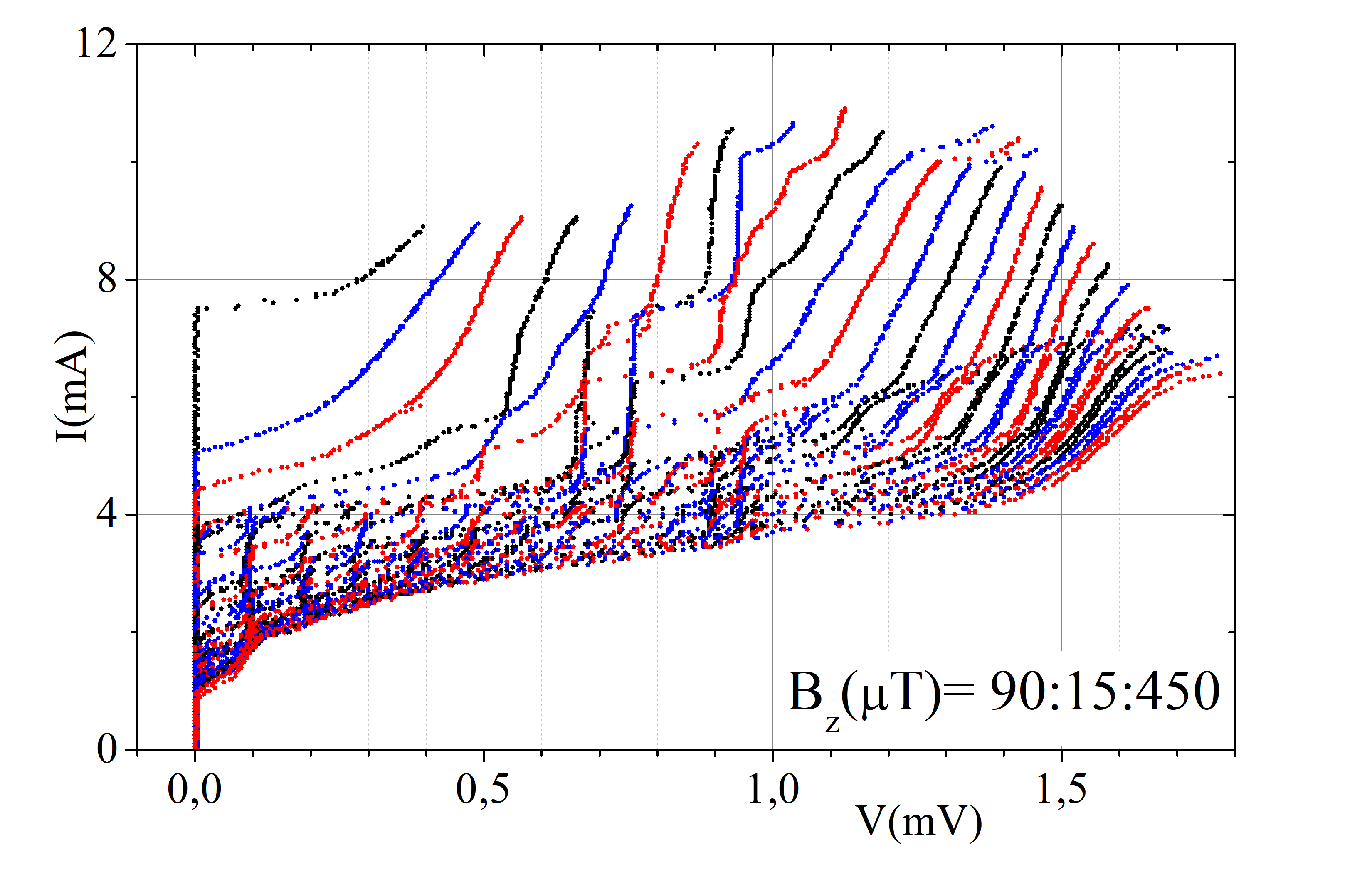}
\caption{(Color online) Family of current-voltage characteristics of the zero-field cooled \conf AJTJ listed in Table I recorded at $4.2\,K$ for different values of a transverse magnetic field, $B_{z}$,  i.e., perpendicular to the junctions plane. $B_z$ ranges from $90\, \mu T$ (the leftmost curve) to $450\, \mu T$ (the rightmost curve) with increments of $15\,\mu T$.}
\label{IVC_ZFC}
\end{figure}

\noindent We now focus on the evolution of the current-voltage characteristics obtained by sweeping the bias current with a triangular waveform on our ZFC \conf AJTJ subject to a gradually increasing transverse field, $B_z$. Fig.~\ref{IVC_ZFC} presents the family of IVCs recorded at $4.2\,K$ at different values of $B_z$, varying from $90\, \mu T$, that is slightly below the transverse critical field, to $450\, \mu T$ in steps of $15\, \mu T$. A sequence of magnetically induced structures at larger and larger voltages, such as displaced linear slopes \cite{Barone71}, Fiske step staircase \cite{Fiske65}, and Eck steps \cite{Eck}, was registered upon increasing the field strength. This succession of current singularities is identical, apart from the very different magnetic field scale, to that stemming from the application of an in-plane field $B_{\bot}$ as investigated in Ref.\cite{PRB19}. The experimental findings of this Section with no trapped field ratify once again that the effects of a transverse magnetic field applied to a Lyngby-type \conf AJTJ are in all respect comparable to those of an in-plane magnetic field applied in the direction of the bias current \cite{SUST15,SUST18}. In other words, the radial distribution of the magnetic field induced by the shielding currents circulating on the outer borders of the top and bottom junction's electrodes is qualitatively similar to that created by an uniform in-plane magnetic field in the $Y$-direction. 
 
\subsection{Field Cooling}

\noindent A large variety of phenomena occurs when a planar \Jos \jun is cooled through its critical temperature in the presence of a magnetic field perpendicular to the its plane. If the junction's electrodes are realized with thin-film type-II superconducting strips, like Nb and high-Tc materials, the magnetic field can be trapped in the form of quantized filaments of flux, or vortices, having a normal core of the size of the superconducting coherence length, $\xi$. In finite-width films the vortices nucleated just below $T_c$, due to thermal activation \cite{Finnemore94}, escape through the edges of the strip; as a result, a complete Meissner expulsion of vortices, substantially independent of the details of pinning and material parameters, is observed below a threshold field, $B_z^*\approx \Phi_0/w^2$, which increases with the decreasing of the film width $w$ \cite{Washington82}. Above this field the Meissner effect is typically incomplete and vortices are trapped with a density increasing approximately linearly with the field amplitude and dependent on the sample defects (such as grain boundaries, normal inclusions, etc.) that may act as a pinning sites \cite{Stan04}. If one or more vortices are trapped in one of the electrodes of a Josephson tunnel junction a significant parallel component of magnetic field can result in the barrier region, causing a local suppression of the junction critical current density \cite{Gubankov92}. Furthermore, as already mentioned, when one of the junction's electrodes has a hole, as in our AJTJs, the magnetic flux can be trapped as a result of fluxoid quantization and conservation in a superconducting ring. For the same reason, some magnetic flux can also be trapped in the form of Josephson vortices, i.e., supercurrent loops in tunnel barrier, also called \textit{fluxons}, as each loop carries one magnetic flux quantum. A fluxon corresponds to a localized $2\pi$-change of the \Jos phase and, as a unique property of topologically closed systems, such as AJTJs, the number of trapped fluxons is conserved and new fluxons can be created only in the form of fluxon-antifluxon pairs. All the above trapping processes are not fully reproducible because additional spontaneous productions of vortices \cite{Polturak03}, fluxoid \cite{PRB09RC} and fluxons \cite{PRB08} occur on a statistical basis  with a probability that increases with the speed of the normal-to-superconducting transition. 

\begin{figure}[b]
\centering
\includegraphics[width=9cm]{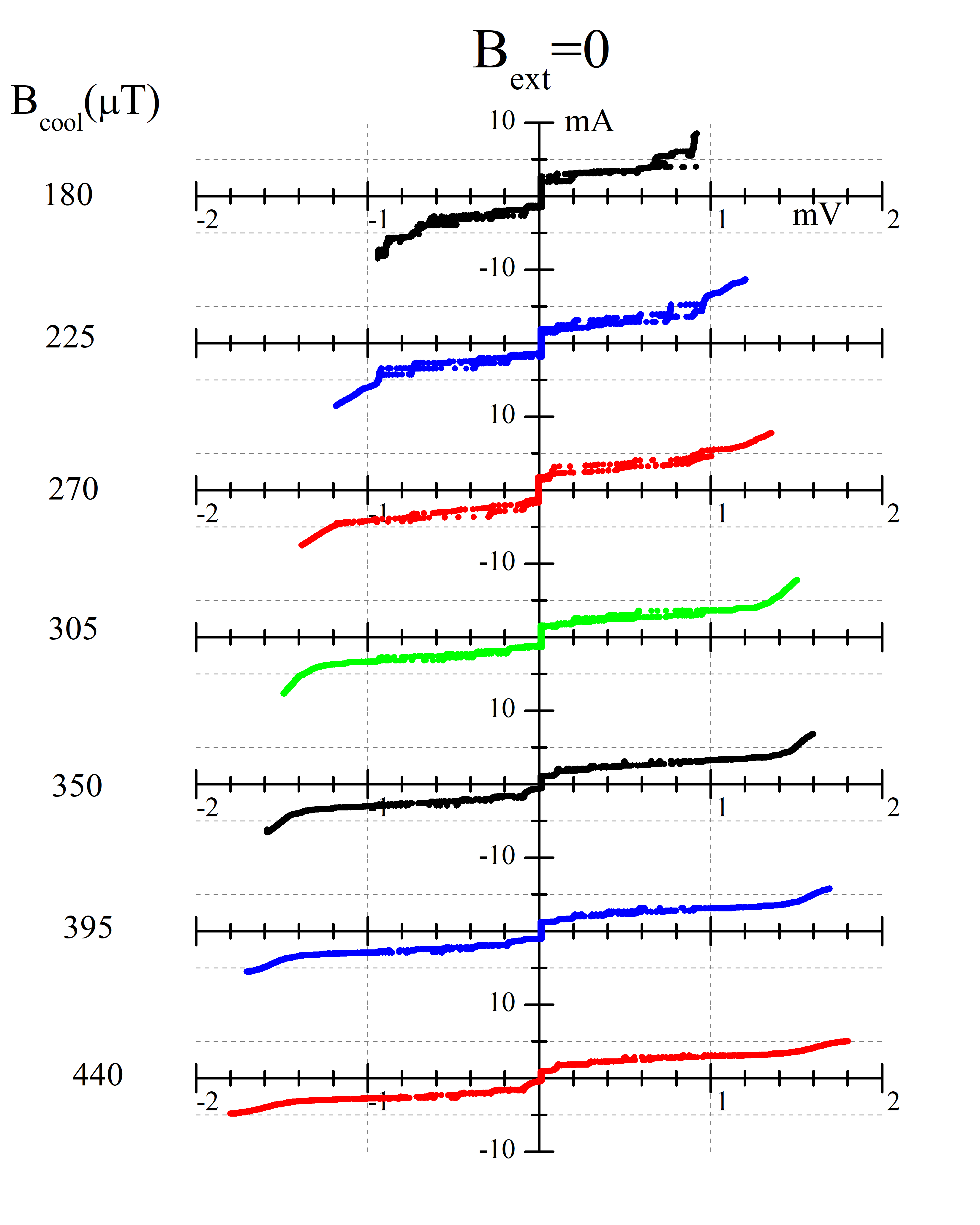}
\caption{(Color online) I-V characteristics of the \conf AJTJ listed in Table I quenched in different transverse fields, $B_{cool}$, whose strength is indicated by the labels. The curves were recorded at $T=4.2\,K$ and in the absence of any externally applied field.}
\label{IVC_FC}
\end{figure}

\noindent Figure~\ref{IVC_FC} shows the IVCs of the \conf AJTJ listed in Table I quenched through its transition temperature down to $T = 4.2\, K$ in the presence of a transverse fields, $B_{cool}$, of increasing amplitudes, as indicated by the labels; after each quench $B_{cool}$ was turned off and the IVC recorded in the absence of any externally applied magnetic field. It is clear that this procedure does not allow a continuous variation of the cooling field; indeed, $B_{cool}$ was changed from 180 to 450$\,\mu T$ in steps of 45$\,\mu T$. In all cases we observe a current singularity whose voltage increases with the cooling field strength; the same pattern is seen reversing the sign of the cooling field. Interestingly, the voltage of each singularity is the same as that of the FFS that would be obtained on the zero-field cooled sample by applying an external transverse field, $B_z$, equal to the corresponding cooling field, $B_{cool}$. However, the height of the current steps are slightly smaller in the case of field quenches. In addition, for each given $B_{cool}$ value, the voltage of the current singularity increases if an external transverse field, $B_z$, is gradually applied with polarity opposite to the cooling field; vice versa, that voltage decreases in the presence of a $B_z$ with the same polarity until the resonance disappears from the IVC and a finite, although small, zero-voltage critical current, $I_c$, is recovered when $B_z$ approaches $B_{cool}$. \textcolor{black}{If the bias is increased beyond the point where the current singularities occur a sudden switch to higher voltage is observed.} According to the observed phenomenology, we classify the branches in Figure~\ref{IVC_FC} as FFSs induced by the persistent current, $I_{circ}$, that circulate in the junction's top electrode to maintain the London fluxoid, $\Phi_f$, trapped in its hole during the field cooling. This current mainly flows in the proximity of the inner perimeter of the superconducting loop \cite{clem03and04} and produces at the ring surface a radial magnetic field, $B_{rad} \propto I_{circ}$. The closer is the inner perimeter to the tunnel barrier, the larger is the effect of the trapped fluxoid. It is well known \cite{Ketchen85} that the effective capture area, $A_{eff}$, of a hole in a superconducting loop is larger than the actual area of the hole, that for our elliptical hole is $A_h=\pi \times 3.5 \mu m \times 38\mu m\approx 420 \mu m^2$. Assuming that $A_{eff}\approx A_h$, the cooling field needed to trap just one magnetic flux quantum is $\Phi_0/A_h \simeq 5\,\mu T$, indicating that a fluxoid, $\Phi_f \propto B_{cool}$, made by several tens of flux quanta is trapped during each quench. The magnetic flux trapped in the hole made in the top electrode must also thread the simply-connected bottom electrode in the form of tens of distributed vortices. As the coherence length of Nb thin-film \cite{Draskovic13} is $\xi^{Nb}\cong 10\,nm$, their interaction range is very small. Therefore, for the large majority vortices the associated persistent currents circulate far from the barrier and their spatially averaged effects is negligible.  


\noindent For the sake of completeness, it must be added that, although care was taken to slowly cool the \jun through its critical current, in few cases quite asymmetric IVCs were observed and, consequently, a new cooling process was attempted without recording. We explain this as due to the random trapping of vortices in the proximity of the tunnel barrier. Also the number of trapped fluxons is neither reproducible nor measurable from one quench to the next. Nevertheless, as a transverse field does not break the symmetry of the \Jos phase, we believe that only few fluxons are spontaneously generated during the quench, if any. 

\begin{figure}[t]
\centering
\includegraphics[width=10cm]{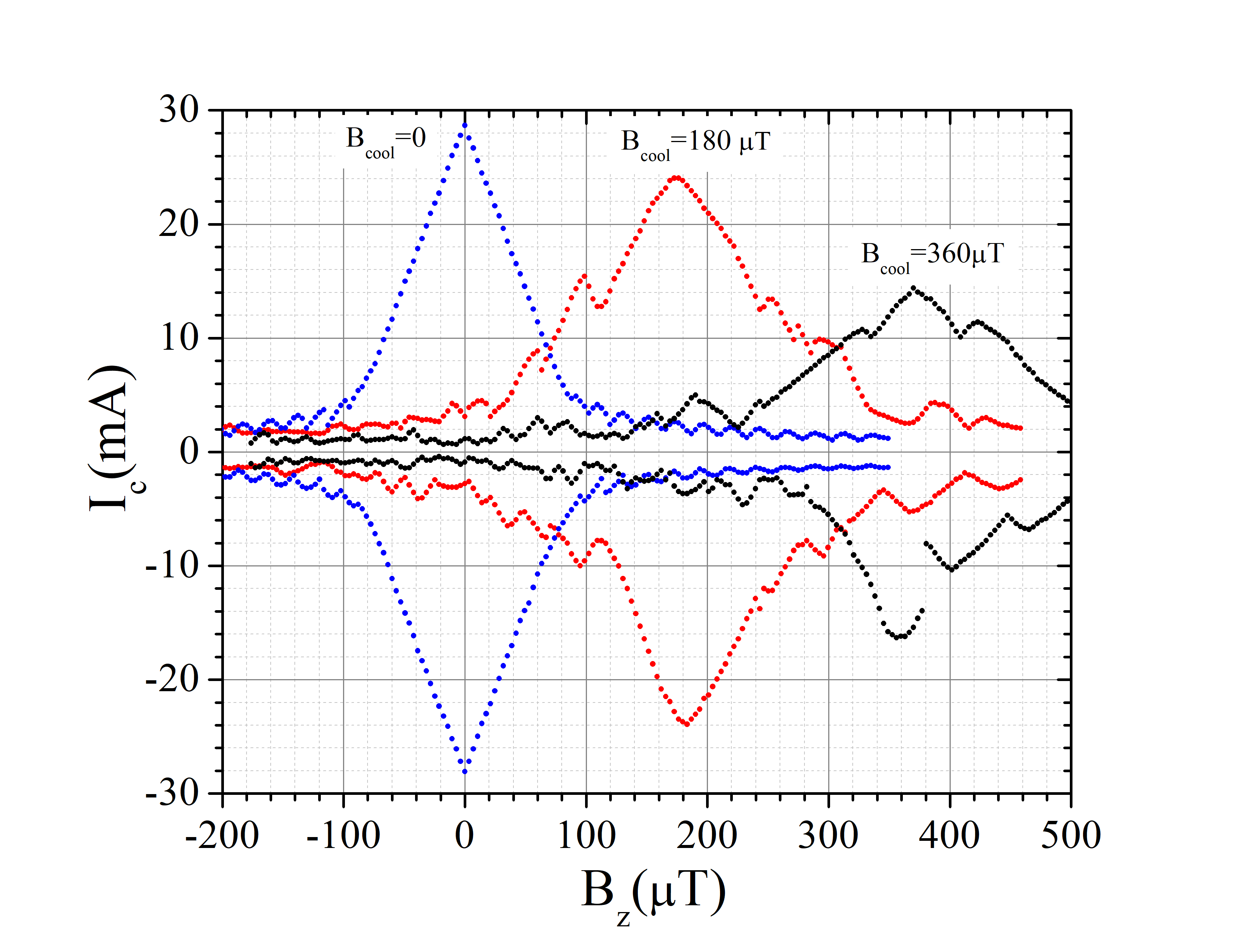}
\caption{(Color online) Transverse magnetic diffraction patterns, $I_c(B_z)$, recorded at $4.2\,K$, after the quench of the \conf AJTJ with different cooling fields.}
\label{MDP_FC}
\end{figure}

\noindent It is interesting at this point to address how the zero-voltage critical current, $I_c$, modulates with an externally applied transverse field, $B_z$, once the junction has been quenched in a cooling field, $B_{cool}$. Figure~\ref{MDP_FC} shows three transverse MDPs, $I_c(B_z)$,: the leftmost threshold curve is the same as that in Figure~\ref{ZFC_MDP}(b) recorded in the flux-free regime ($B_{cool}=0$). The interference pattern in the middle has been recorded after a quench at $B_{cool}=180\,\mu T$ and the rightmost curve was obtained after a quench at $B_{cool}=360\,\mu T$. Although the field-cooled MDPs are not at all reproducible from one quench to another, they show the common characteristic to have their maximum where the applied transverse field is approximately equal to the cooling field. In addition, the MDP largest value decreases with the increasing cooling field. We explain the progressive degradation and the loss of symmetry of the $I_c(B_z)$ curves as due to the increasing density of randomly trapped vortices nearby the tunnel barrier and in both junction electrodes. In fact, being the bottom electrode $100\,\mu m$ wide, its threshold value for the vortices trapping, $\Phi_0/w^2\approx 0.2\,\mu T$, is way smaller than our cooling field values. There is also another important conclusion that can be drawn from the fact that the smallest $I_c$ modulation occurs for $B_z\simeq B_{cool}$; at the very end of each quench, before removing the cooling field, the system is in what some authors \cite{Ketchen85,clem03and04} called \textit{flux focusing state} in which the net circulating current in an isolated superconducting ring is null, i.e., the persistent currents flowing in the electrode interior are balanced by the shielding current flowing on the electrode border in opposite direction. More strictly, in our case we can state that the net currents are smallest when $B_z\simeq B_{cool}$. 

\begin{figure}[t]
\centering
\includegraphics[width=8cm]{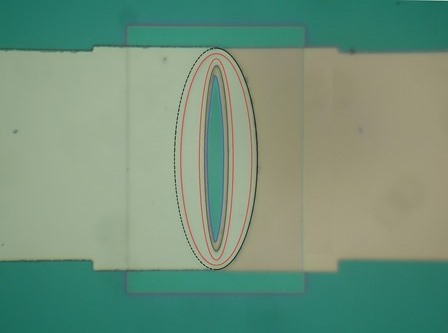}
\caption{(Color online) Optical image of a \textit{Lyngby-type} confocal annular Josephson tunnel junction made by the superposition of two $Nb$ doubly-connected electrodes. For this sample the ratio of the minor axis and the major axis is $1\!:\!4$ which implies that the equatorial annulus width is one forth of the polar width.}
\label{picture}
\end{figure}



\noindent Interestingly, experimental findings quantitatively indistinguishable from those reported so far were obtained in samples, as that shown by the optical image of Figure~\ref{picture}, in which both electrodes are doubly connected. In different words, during the field cooling of a AJTJ, it is irrelevant whether or not also the base electrode has a hole. \textcolor{black}{It is also worth noting that the data reported in this section show that variation of the magnetic field of few tens of microteslas drastically affect the junction's I-V characteristic. Therefore, if the experiments were carried out without shielding the Earth's magnetic field, a considerable magnetic shift would have been observed.} 


\section{The modeling} 

A theoretical interpretation of the flux-flow state observed in field-cooled AJTJs will be given in the this section. The perturbed sine-Gordon equation has always been the most adequate phenomenological model to describe the electrodynamics of long JTJs in the presence of bias current, magnetic fields and losses \cite{Barone}. The geometry of our AJTJs suggests the use of the (planar) elliptic coordinate system $(\nu,\tau)$, a two-dimensional orthogonal coordinate system in which the coordinate lines are confocal ellipses and hyperbolae. Upon assuming that the \conf annulus is narrow, $\Delta w_{max}<< \lambda_J$, the \Jos phase, $\phi$, does not depends on the radial coordinate, $\nu$, and the system becomes one-dimensional, that is, the spatial dependence of $\phi$ is only determined by the angular elliptic coordinate, $-\pi \leq \tau \leq \pi$, that reduces to the angular polar coordinate in a circularly symmetric system. In our notations the origin of $\tau$ coincides with the $Y$-axis in Figure~\ref{ConfAnn} and increases with a clockwise rotation. It was derived that a \conf AJTJ in the presence of an externally applied spatially homogeneous in-plane magnetic field, ${\bf H^{ext}}$, of arbitrary orientation, $\bar{\theta}$, relative to the $Y$-axis, obeys a modified and perturbed sine-Gordon equation \cite{JLTP16b}:

\begin{equation}
 \left[\frac{\lambda_J}{c\,\mathcal{Q}(\tau)}\right]^2 \left(1+\beta\frac{\partial}{\partial \hat{t}}\right) \phi_{\tau\tau} - \phi_{\hat{t}\hat{t}}-\sin \phi =\alpha \phi_{\hat{t}} - \gamma(\tau) + F_h^{ext}(\tau),
\label{psge}
\end{equation}

\noindent where $\hat{t}$ is the time normalized to the inverse of the so-called (maximum) plasma frequency, $\omega_p^{-1}=\sqrt{\Phi_0 c_s/2\pi J_c}$ (with $c_s$ the specific junction capacitance) and the critical current density, $J_c$, was assumed to be constant. Here and in the following, the subscripts on $\phi$ are a shorthand for derivative with respect to the corresponding variable. $\mathcal{Q}(\tau)$ is the elliptic scale factor defined by $\mathcal{Q}^2(\tau) \equiv \sinh^2\bar{\nu} \sin^2\tau+\cosh^2 \bar{\nu} \cos^2 \tau= \sinh^2\bar{\nu}+ \cos^2\tau=\cosh^2\bar{\nu} - \sin^2\tau=(\cosh2\bar{\nu} + \cos2\tau)/2$, where $\bar{\nu} \equiv \arctanh \rho$ is an alternative measure of the annulus eccentricity, $e^2 \equiv 1-\rho^2=\sech^2\bar{\nu}\leq 1$. Furthermore, $\gamma(\tau)\equiv J_Z(\tau)/J_c$ is the normalized bias current density and 

\begin{equation}
F_h^{ext}(\tau)\equiv h^{ext} \frac{\cos\bar{\theta}\cosh\bar{\nu} \sin\tau-\sin\bar{\theta}\sinh\bar{\nu}\cos\tau }{\mathcal{Q}^2(\tau)}
\label{Fh}
\end{equation}
\noindent is an additional forcing term proportional to the in-plane applied magnetic field; $h^{ext}\equiv H^{ext}/J_c c$ is the normalized field strength for treating long \conf AJTJs. As usual, the $\alpha$ and $\beta$ terms in Eq.(\ref{psge}) account for, respectively, the quasi-particle shunt loss and the surface losses in the superconducting electrodes. Eq.(\ref{psge}) is supplemented by the periodic boundary conditions \cite{PRB96}:

\begin{subequations}
\begin{eqnarray} 
\label{peri1}
\phi(\tau+2\pi,\hat{t})=\phi(\tau,\hat{t})+ 2\pi n_w,\\
\phi_\tau(\tau+2\pi,\hat{t})=\phi_\tau(\tau,\hat{t}),
\label{peri2}
\end{eqnarray}
\end{subequations}

\noindent where the integer $n_w$, called the \textit{winding number}, is the algebraic sum of the flux quanta trapped in each electrode when cooled below its critical temperature and counts the number of fluxons trapped in the \jun barrier. Eq.(\ref{psge}) can be classified as a perturbed and modified sine-Gordon equation in which the perturbations are given by the system dissipation and driving fields, while the modification is represented by an effective local $\pi$-periodic \Jos penetration length, $\Lambda_J(\tau)\equiv \lambda_J/Q(\tau)= c \lambda_J \Delta \nu /\Delta w(\tau)$, inversely proportional to the annulus width, $\Delta w(\tau)\equiv\mathcal{Q}(\tau)\,\Delta\nu$. 


\subsection{The Effect of the Trapped Fluxoid} 

The forcing term in Eq.(\ref{Fh}) has been very successfully used in Ref.[1] to numerically reproduced the evolution of the current singularities induced in \conf AJTJs by an in-plane magnetic field. According to the argumentation in the previous Section, $F_h^{ext}$ would equally well reproduce the FFSs induced in zero-field cooled AJTJs by a transverse magnetic field as reported in Figure~\ref{IVC_ZFC}. We now want to find the forcing term, $F_h^{rad}(\tau)$, that takes into account the radial field, $H^{rad}$, generated by the persistent current circulating in the inner perimeter of the hole in the top junction's electrode when quenched in a transverse magnetic field (that is removed once the temperature is well below the critical temperature). For this purpose it is convenient to resort to the general equation of motion for the \Jos phase developed by Goldobin \textit{et al.} \cite{Goldobin01} for one-dimensional curved variable-width JTJs in the presence of an arbitrary externally applied in-plane magnetic field, ${\bf {H}}$. According to this theory, although adopting our notations, $\phi(\hat{s},\hat{t})$ satisfies the following non-linear PDE:

\begin{equation}
\phi_{\hat{s}\hat{s}} - \phi_{\hat{t}\hat{t}}-\sin \phi =\gamma+ \alpha \phi_{\hat{t}} + \frac{1}{J_c \lambda_J} \frac{dH_\nu}{d\hat{s}}+ \frac{\Delta w_{\hat{s}}}{\Delta w}\left[\frac{H_\nu}{J_c \lambda_J}-\phi_{\hat{s}}\right],
\label{goldo}
\end{equation}

\noindent where $\hat{s}=s/\lambda_J$ is the normalized curvilinear coordinate (for the sake of simplicity, the surface losses were neglected in Ref.\cite{Goldobin01}). $H_\nu(\tau)\equiv{\bf {H}}\cdot{\bf \hat{N}}$, with ${\bf \hat{N}}$ being the the (outward) normal unit vector to the \conf annulus, is the component of the applied magnetic field normal to the junction perimeter. $\Delta w_s$ is the directional derivative of the local \jun width, $\Delta w$. Making use of the equality \cite{JPCM16}:

\begin{equation}
\nonumber
\frac{d^2}{d\hat{s}^2}+ \frac{\Delta w_{\hat{s}}}{\Delta w} \frac{d}{d\hat{s}} \equiv \left(\frac{\lambda_J}{c \mathcal{Q}}\right)^2\frac{d^2}{d\tau^2},
\label{no1}
\end{equation}

\noindent Eq.(\ref{goldo}) can be rearranged as:

\begin{equation}
\left[\frac{\lambda_J}{c\,\mathcal{Q}(\tau)}\right]^2 \phi_{\tau\tau}  - \phi_{\hat{t}\hat{t}}-\sin \phi =\gamma+ \alpha \phi_{\hat{t}} + \frac{1}{J_c \lambda_J} \left[ \frac{dH_\nu}{d\hat{s}}+ \frac{\Delta w_{\hat{s}}}{\Delta w} H_\nu\right].
\label{goldo2}
\end{equation}

\noindent It has been shown \cite{JPCM16} that Eq.(\ref{goldo2}) reduces to Eq.(\ref{psge}), if ${\bf H}$ is a uniform in-plane magnetic field, ${\bf H^{ext}}$. When, as in our case, ${\bf H}$ is a in-plane field with a constant radial component $H_\nu(\tau)=H^{rad}$, then, recalling that $ds=c \mathcal{Q}(\tau) d\tau$ and exploiting the fact that, in elliptic coordinates, $\Delta w_{\hat{s}}/\Delta w =-\lambda_J \sin2\tau/2c \mathcal{Q}^3(\tau)$, after some algebraic manipulations, we get: 
\begin{equation}
\nonumber
\frac{dH^{rad}}{d\hat{s}} + \frac{\Delta w_{\hat{s}}}{\Delta w} H^{rad} = \frac{\lambda_J}{c \mathcal{Q}} \frac{dH^{rad}}{d\hat{\tau}} - \frac{\lambda_J}{2 c} H^{rad} \frac{\sin 2\tau}{ \mathcal{Q}^3}=- \frac{\lambda_J}{2 c} H^{rad} \frac{\sin 2\tau}{ \mathcal{Q}^3}.
\label{no2}
\end{equation}

\noindent Therefore, by inserting the last expression into Eq.(\ref{goldo2}),	we get the new magnetic forcing term in Eq.(\ref{psge}) due to a trapped fluxoid, namely:

\begin{equation}
F_h^{rad}(\tau) = \frac{1}{J_c \lambda_J }\left[\frac{dH^{rad}}{d\hat{s}} + \frac{\Delta w_{\hat{s}}}{\Delta w} H^{rad}\right] = - \frac{H^{rad}}{J_c c}  \frac{\sin 2\tau}{2 \mathcal{Q}^3}= - h^{rad}  \frac{\sin 2\tau}{2 \mathcal{Q}^3},
\label{radialforcing}
\end{equation}

\noindent with $h^{rad}\equiv H^{rad}/J_c c$ proportional to the circulating current, $I_{circ}$, that, in turn, is proportional to the trapped fluxoid, $\Phi_f$. By replacing $F_h^{ext}$ with $F_h^{rad}$ in Eq.(\ref{psge}) we can model a field-cooled AJTJ in the absence of any external magnetic field, even though we do not know the constant of proportionality between $h^{rad}$ and $\Phi_f$ (and the cooling field). It is readily seen that $F_h^{ext}(\tau)$ and $F_h^{rad}(\tau)$ have different spatial periodicity, as the former is $2\pi$-periodic, while the latter is $\pi$-periodic. It should be noted, in addition, that the radial forcing term vanishes as the junction's aspect ratio tends to unity (as $\rho \longrightarrow 1$, $\bar{\nu} \longrightarrow \infty$); in different words, no effect of a trapped fluxoid can be observed in a circular AJTJ as its barrier has a constant width. 


\subsection{Numerical simulations}

The commercial finite element simulation package COMSOL MULTIPHYSICS (www.comsol.com) was used to numerically solve Eq.(\ref{psge}) subjected to the cyclic boundary conditions in Eqs.(\ref{peri1}) and (\ref{peri2}). In order to compare the numerical results with the experimental findings presented in the previous section, we set the annulus normalized length, $\ell=L/\lambda_J=50$ and aspect ratio, $\rho=1/4$. We have assumed a uniform current distribution, i.e., $\gamma(\tau)= \gamma_0$. The damping coefficient $\alpha$ was changed in the weakly underdamped region $0.1 \leq\alpha\leq 0.3$, while the surface losses were simply neglected ($\beta=0$) to save computer time. Throughout this section we will assume that no 
fluxons were trapped in the barrier at the time of the normal-to-superconducting transition, i.e., we set the winding number, $n_w$, to zero in the periodic condition of Eq.(\ref{peri1}). 

\begin{figure}[t]
\centering
\includegraphics[width=8cm]{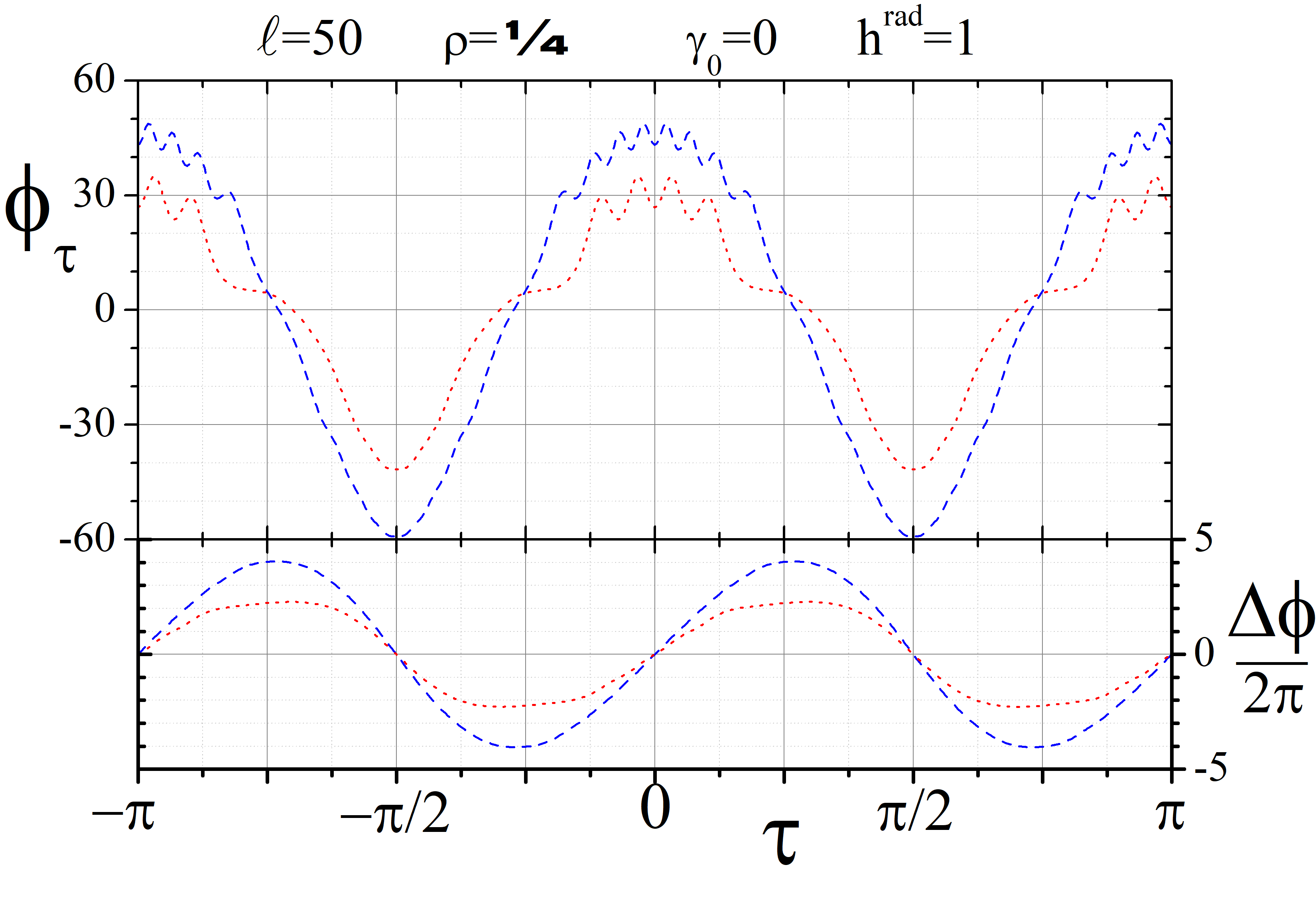}
\caption{(Color online) Two static numerical solutions of Eq.(\ref{psge}) obtained for $\ell=50$, $\rho=1/4$, $\gamma_0=0$ and $h_{rad}=1$. Bottom panel: phase variation, $\Delta \phi(\tau) \equiv \phi(\tau) - \phi(0)$, normalized to $2\pi$ (see right vertical scale). Top panel: phase spatial derivative, $\phi_\tau$. See text.}
\label{staticprofiles}
\end{figure}

\noindent We begin the numerical investigation by searching the static, i.e, time-independent solutions obtained in the absence of a bias current ($\gamma_0=0$) for different values of the radial field, $h^{rad}$, in Eq.(\ref{radialforcing}). It was found that for a given radial field, depending on the initial condition, several static profiles, $\phi(\tau)$, satisfy the PDE which differ by the number of the occurring 2$\pi$-kinks; these multiple static solutions are typical of long JTJs in the presence of an external magnetic field \cite{owen}. Clearly, for a AJTJ with no initially trapped fluxons, the number of positive kinks (fluxons) must match exactly that of negative kinks (antifluxons) in order to have a single-valued periodic \Jos phase. Figure~\ref{staticprofiles} shows two of the several static solutions existing for $h^{rad}=1$; the bottom panel concerns the phase variation, $\Delta \phi(\tau) \equiv \phi(\tau) - \phi(0)$, normalized to the kink size, $2\pi$, while the top panel shows the phase spatial derivatives, $\phi_\tau$. From the bottom panel we notice that the phase profiles have a $\pi$-periodicity with minima in $-\pi/4$ and $3\pi/4$ and maxima in $-3\pi/4$ and $\pi/4$; at the first order, they can be approximated by a $\sin 2\tau$ function. The two solutions shown in Figure~\ref{staticprofiles} mainly differ by their amplitudes: the dotted line corresponds to a phase swing of $4.6 \times 2\pi$, while the phase variation of dashed curve is $8.0 \times 2\pi$. In different words, for the dotted (dashed) solution a static chain of between four and five (eight) fluxons exists between each phase minimum and its adjacent maximum (positive $\phi_{\tau}$) and the same number of antifluxon make up the chain standing between each phase maximum and its nearest minimum (negative $\phi_{\tau}$). This can be seen looking at the phase derivatives shown in the top panel of the figure, although a lack of symmetry is evident between the positive and negative parts. In fact, the positive $\phi_{\tau}$ peaks associated with the fluxons are well resolved while the negative peaks related to the antifluxons are smeared out. This is ascribed to the fluxon repelling (attracting) barrier induced by a widening (narrowing) \Jos transmission line \cite{nappipagano} that makes the physics of \conf AJTJs very rich and interesting. As the barrier polarity is the same for fluxons and antifluxons \cite{JLTP16b}, the fluxon repel each other at the polar points ($\tau=0$ and $\pm \pi$) where the annulus is widest, while the antifluxons attract each other as they are gathered at the equatorial points ($\tau=0$ and $\pm \pi$) where the annulus width is smallest. Our simulations showed that the number of existing static solutions increase with the amplitude of the radial field. Furthermore, the quantity of kinks grouped in, either positive or negative, static chains grows continuously with $h^{rad}$ (but never exceeds $10 h^{rad}$). Therefore, as an example, for $h^{rad}=1$ we can have as many as 40 static kinks in the phase profile. 

\noindent When a dc bias current is applied to the AJTJ, both the fluxons and antifluxons experience a Lorentz force the direction of which depends on their polarity. For a bias sufficiently large to overcome the static friction the two chains of fluxons and the two chains of antifluxons get depinned from the potential wells and start to move in opposite direction. It is the motion of the fluxons and antifluxons that sets the junction in the finite voltage state. The motion of a single fluxon along a \conf AJTJ is non-uniform and, due to both the tangential and radial acceleration, plasma waves are emitted by the leading (trailing) edge of the accelerating (decelerating) fluxon. When, as in our case, dense fluxon trains collide with dense antifluxon trains, a wealth of wide-spectrum radiation is generated. It is not surprising that, due to the many internal degrees of freedom in the moving fluxon chains, quasi-periodic or chaotic dynamic solutions are quite often obtained when numerically solving the perturbed sine-Gordon equation \cite{Ustino96}. Recently a chaotic system based on an extended JTJ has been also proposed as a withe-noise source in the terahertz region \cite{Koshelets19}. Indeed the parameter space where periodic solutions exist is limited and even more restricted is the region where large-amplitude resonances are observed. 

\begin{figure}[t]
\centering
\subfigure[ ]{\includegraphics[width=8cm]{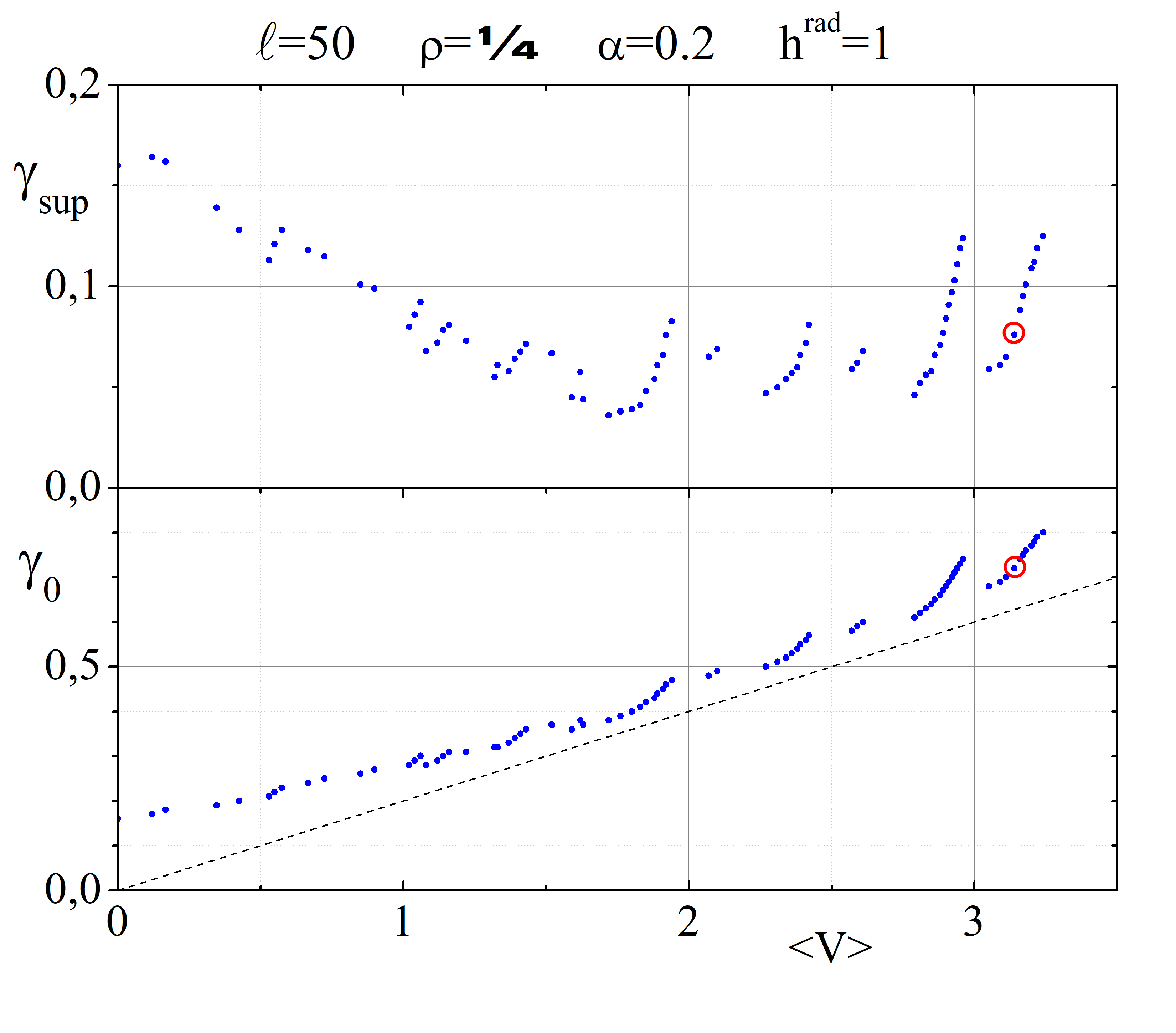}}
\caption{(Color online) Bottom panel: Numerically computed current-voltage characteristics of a \conf AJTJ with aspect ratio $1/4$ and normalized length $\ell=50$ obtained by fixing the loss parameter $\alpha=0.2$ and the normalized radial magnetic field $h_{rad}=1$. The dotted lines indicate the ohmic current, $\gamma_{nor}=\alpha <\!\!V\!\!>$. Top panel: as in the bottom panel but with the background ohmic current subtracted, $\gamma_{sup}\equiv \gamma_0-\gamma_{nor}$. }
\label{IVCnum}
\end{figure}

\noindent The bottom panel of Fig.~\ref{IVCnum} shows the numerically computed current-voltage characteristics, $\gamma_0$ vs $<\!\!V\!\!>$, obtained for $h^{rad}=1$ by fixing the loss parameter $\alpha=0.2$ and starting the calculation with a static solution consisting of fluxon chains each made by about 7 fluxons (or antifluxons). The dotted lines indicate the ohmic current, $\gamma_{nor}=\alpha <\!\!V\!\!>$. Each point in the plots corresponds to a flux-flow dynamical state whose time evolution will be considered below. Such solutions are periodic in time and space and their frequency, $2\pi/T$, with $T$ being the time periodicity, is identified by the normalized average voltage, $<\!\!V\!\!>$, that could also be evaluated by averaging $\phi_{\hat{t}}(\tau,\hat{t})$ over a sufficiently long time. It is seen that the numerically computed FFSs consists of a set of almost equally spaced high-order Fiske steps whose voltage position increases with the field strength. The voltage width of each Fiske resonance is approximately equal to $\alpha$ while the voltage separation $\Delta\!<\!\!V\!\!>$ between two adjacent steps is about $0.24$, i.e., twice the voltage separation reported for Fiske steps computed for a \conf AJTJ with the same geometry but in a uniform in-plane field \cite{PRB19}. We believe that this can be ascribed to the halved periodicity of the radial magnetic forcing term $F_h^{rad}$. This effect is better evidenced in the top panel of Fig.~\ref{IVCnum} where the same data are replotted in terms of the supercurrent, $\gamma_{sup}\equiv\gamma_0-\gamma_{nor}$, which is computed as the spatio-temporal average of $\sin \phi(\tau,\hat{t})$ and provides a measure of the stability of the dynamical state. 

\begin{figure}[t]
\centering
\includegraphics[width=8cm]{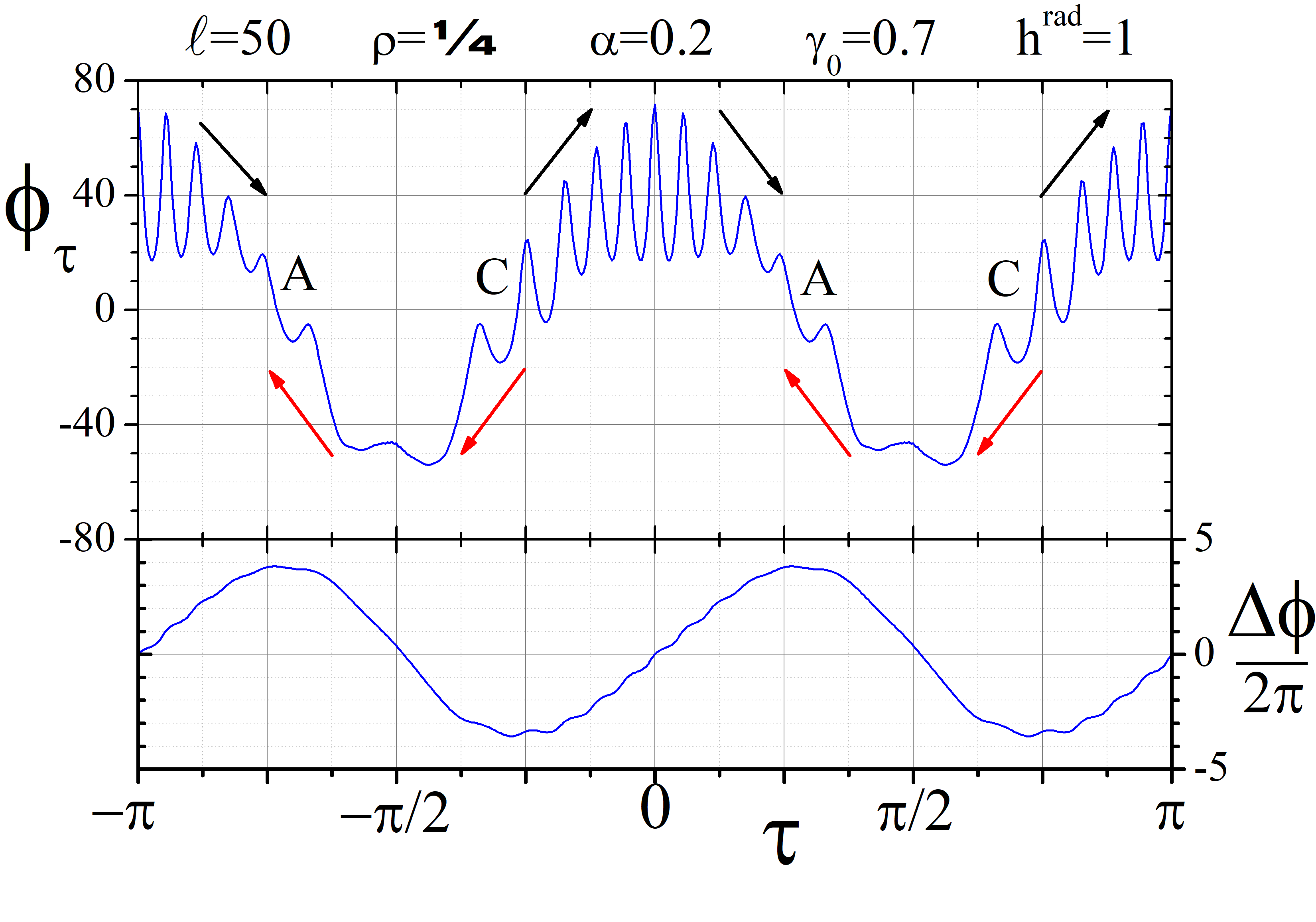}
\caption{(Color online) Numerically computed phase profile (bottom panel) and its spatial derivative (top panel) obtained for $\rho=1/4$, $\ell=50$, $\alpha=0.2$, $\gamma_0=0.7$ and $h_{rad}=1$ that corresponds to the point marked by the open circle in Fig.~\ref{IVCnum}.}
\label{ux}
\end{figure}

\noindent The time evolution of the numerical solutions of Eq.(\ref{psge}) is qualitatively illustrated in Fig.~\ref{ux} which shows the phase profile (bottom panel) and is spatial derivative (top panel), taken at an arbitrary time and computed for $\rho=1/4$, $\ell=50$, $\alpha=0.2$, $\gamma_0=1$ and $h_{rad}=1$, which corresponds to the point marked by an open circle in Fig.~\ref{IVCnum}. In the presence of a radial field, fluxon-antifluxon pairs are continuously created at the points, pinpointed by the letter $C$, where the phase is smallest. Under the influence of the Lorentz forces due to the bias current and the magnetic field, the fluxons (positive pulses) circulate clockwise (increasing $\tau$), as indicated by the black arrows, while the antifluxons (negative pulses) rotate anticlockwise (decreasing $\tau$), as indicated by the red arrows. Since, they travel with opposite but equal speed, they collide and annihilate at the diametrically opposite points, identified by the letter $A$, corresponding to phase maxima. In passing, we recall that the sign of the magnetic potential felt by a fluxon depends on its polarity. In the presence of dissipative effects, two colliding fluxon and antifluxon fully annihilate if their velocity is below a threshold that increases with the losses \cite{Scott78}. Therefore, by increasing the bias current a speed is reached where the kinks pass through each other without mutual destruction. When dense trains of fluxons collide the situation is more complicated as the leading kinks may exit the first collision with a reduced speed and the fading out during the second collision, and so on. However, as the number of collisions increases the growing radiation makes the system unstable and the Josephson phase suddenly switches to a uniformly rotating profile characterized by a very large voltage. It turned out that the complete trains annihilation is the necessary requirement for a periodic dynamical solution and a stable flux-flow process. Otherwise, the system admits either chaotic or trivial solutions. The eccentricity of the \conf AJTJ plays a determinant role in our self-sustained \Jos flux-flow; in fact, as the confocal annulus tends to a circular ring, the potential well disappears and the fluxon-antifluxon annihilation becomes less likely.

\section{Comments and conclusions}

We have modeled a field cooled AJTJ considering the effect of the fluxoid trapped in the hole made in just one of the electrode forming the junction (the top electrode in our samples). The resulting magnetic forcing term in Eq.(\ref{radialforcing}) is proportional, through $h_{rad}$, to the persistent current circulating in the proximity of the hole perimeter. Clearly, the effective radial field felt by the tunnel barrier depends on how close this perimeter is to the tunnel barrier. When both electrodes are doubly connected, the persistent currents circulate in the inner perimeter of both superconducting holes. As these currents flow on the opposite sides of the barrier, their radial fields have a opposite signs and tend to cancel each other. However, for technical reason the two holes, although concentric, do not have the same area; more specifically, as Figure~\ref{picture} shows, the hole in the bottom electrode is considerably smaller than that in top electrode, that is, its inner perimeter runs far away form the barrier. This asymmetry makes the radial field generated in the top electrode dominant and explains why the experimental findings are much the same in AJTJs with just one or both doubly connected electrodes.

\begin{figure}[t]
\centering
\includegraphics[width=9cm]{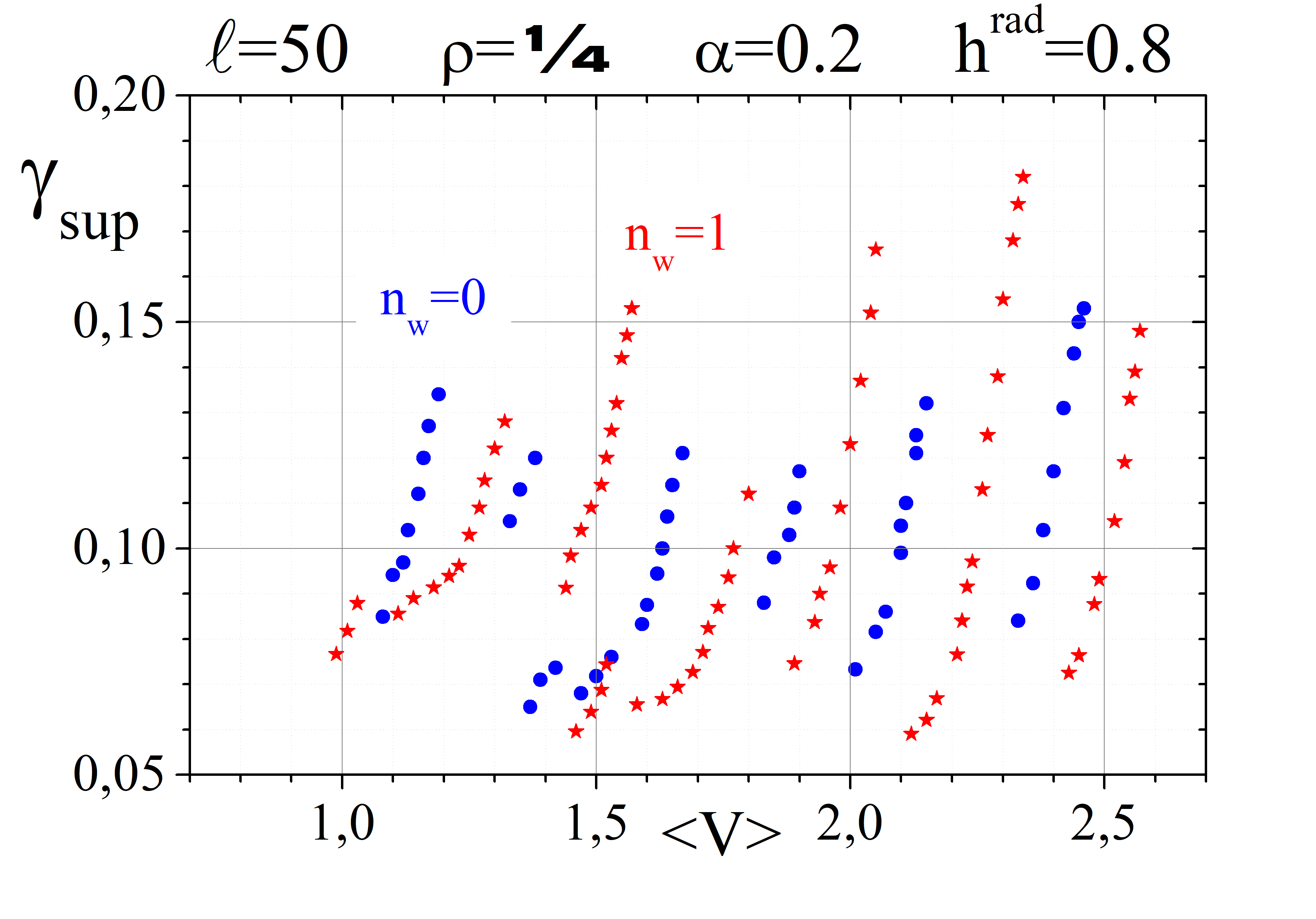}
\caption{(Color online) Comparison of the numerical supercurrent-voltage characteristics computed for a \conf AJTJ without (circles) and with one (stars) trapped fluxon. The calculations were carried out for $\ell=50$, $\rho=1/4$, $\alpha=0.2$ and $h_{rad}=0.8$.}
\label{comparison}
\end{figure}

\noindent So far, in our analysis we have neglected the effects of the vortices trapped in the superconducting films mainly because they cannot be taken into account in the perturbed sine-Gordon equation governing the system. Yet, experiments showed that, although these vortices drastically affect the magnetic dependence of the zero-field critical current, their presence cannot be evidenced from the magnetically induced current singularities. We also considered, for simplicity, that no fluxons are trapped in the barrier at the time of the superconducting quench. This assumption is not realistic and very likely one or more \Jos vortices are trapped during a non-adiabatic quench. Luckily, the numerical analysis can be carried out for an arbitrary number of initially trapped fluxons by changing the winding number, $n_w$, in the periodic conditions Eq.(\ref{peri1}). In Figure~\ref{comparison} we show the numerical supercurrent-voltage characteristics computed for a \conf AJTJ without (circles) and with one (stars) trapped fluxon. We see that the main effect of one trapped fluxon is a voltage shift of the Fiske steps equal to about one half of their voltage separation. As about twelve fluxons-antifluxons participate in the dynamical state it is not surprising that the presence of one extra kink results only in a small relative voltage change.

\vskip 18pt
\noindent In conclusion, despite the many ways that the magnetic flux can be trapped in a planar \Jos tunnel \jun that crosses its critical temperature in the presence of an external magnetic field, as far as we regard \ann junctions, the main effect of the field cooling is due the fluxoid trapped in the hole made in top superconducting electrode. We have considered cooling induction fields perpendicular to the junction plane with amplitude in the microtesla range that is large enough to trap vortices in the thin-films, but way too small to exhibit hysteresis in their magnetization curves. If we had chosen cooling fields sufficiently weak to guarantee the complete Meissner expulsion of the vortices from the films, the trapped fluxoid would have been too small to sustain a \Jos flux-flow and only a small modulation of the \jun zero-voltage critical current would have been observed as a result of the FC process. Experiments on under-damped $Nb/Al$-$AlOx/Nb$ \conf AJTJs showed that, once the cooling field is removed, the magnetic field associated with the conserved fluxoid manifest itself as large-voltage current singularity in the \jun current-voltage characteristics which does not require the application of an external magnetic field. Both the randomly trapped vortices in the electrodes and the randomly trapped fluxons in the tunnel barrier, for different reasons, play a marginal role. Numerical simulations carried on a perturbed sine-Gordon equation, devised to take into account the radial field of the trapped fluxoid, demonstrate that the magnetic resonances correspond to complicated kinks dynamical states consisting of two diametrically opposite trains of fluxons that move towards two diametrically opposed trains of antifluxons. The key ingredient of this dynamics is the $\pi$-periodic magnetic potential established by the persistent current. This potential depends on some geometrical details of the junction. In fact, its amplitude increases as the perimeter of the superconducting hole runs closer and closer to inner barrier boundary. However it vanishes for circular \ann \juns which have a unitary aspect ratio. In different words, the more eccentric is the annulus, the stronger is the influence of the trapped fluxoid. 
 
\noindent Our numerical investigation reproduce, at least at a qualitative level, most of the features of the magnetically-induced steps, such as their profile and field-dependent voltage position. Nevertheless, the step amplitudes and, more generally, the region of stability in the parameters space are larger in the experimental findings. We believe, that due to any even small error in the mask alignment, the annular barrier and the hole in the top electrode are not perfectly concentric, as can be inferred by a careful look at Figure~\ref{picture}: this implies that the persistent current flows at a variable distance from the tunnel barrier. Therefore, an extra $2\pi$-periodic magnetic forcing should be added in the perturbed sine-Gordon Eq.(\ref{psge}) to make a more realistic modeling. 
 
\section*{Acknowledgments}
\noindent RM acknowledges the support from the Italian CNR under the Short Term Mobility Program 2018. RM and JM acknowledge the support from the Danish Council for Strategic Research under the program EXMAD. VPK acknowledges the support from the Russian Foundation for Basic Research, grant No 19-52-80023. The fabrication of the tunnel circuits was carried out by using USU 352529 facilities at the Kotel'nikov IREE RAS within the framework of the state task.

\newpage

\newpage


\begin{thebibliography}{99}
\bibitem{note1} As the ellipse has two axes of symmetry, it is expected that the response of a CAJTJ to the in-plane magnetic field is strongest when the magnetic field is perpendicular to the major axes, as it occurs in elliptical JTJs \cite{Peterson}. It has been reported that for \conf AJTJS the in-plane magnetic diffraction patterns of the zero-voltage critical current, $I_c(H)$, obtained with a field perpendicular, $H_{\bot}$, and parallel, $H_{\parallel}$ to the major axis, differ from one another not only quantitatively but also qualitatively \cite{SUST15}.
\end{thebibliography}
\end{document}